\documentclass[prl,twocolumn,amsmath,showpacs,floats]{revtex4}
\usepackage{graphics}
\usepackage[dvipdfm]{graphicx}
\usepackage{epsfig}
\usepackage{dcolumn}
\usepackage{amsmath}

\begin{document}

\title{A stacking-fault based microscopic model for platelets in diamond}

\author{C. R. Miranda}
\altaffiliation[Current address: ]{The Abdus Salam International
Centre for Theoretical Physics, Strada Costiera 11, 34100, Trieste
- Italy} \affiliation{Instituto de F\'{\i}sica ``Gleb Wataghin'',
Universidade Estadual de Campinas,\\
CP 6165, CEP 13083-970, Campinas, SP, Brazil}

\author{A. Antonelli}
\affiliation{Instituto de F\'{\i}sica ``Gleb Wataghin'',
Universidade Estadual de Campinas,\\
CP 6165, CEP 13083-970, Campinas, SP, Brazil}
%\address{$^{(1)}$ Instituto de F\'{\i}sica ``Gleb Wataghin'',
%Universidade Estadual de Campinas,\\
%CP 6165, CEP 13083-970, Campinas, SP, Brazil}

\author{R. W. Nunes}
\affiliation{Departamento de F\'{\i}sica, Universidade Federal de
Minas Gerais, Belo Horizonte, Minas Gerais, CEP 30123-970, Brazil}

\date{\today}

\begin{abstract}

We propose a new microscopic model for the $\{001\}$ planar
defects in diamond commonly called platelets. This model is based
on the formation of a metastable stacking fault, which can occur
because of the ability of carbon to stabilize in different bonding
configurations. In our model the core of the planar defect is
basically a double layer of three-fold coordinated $sp^2$ carbon
atoms embedded in the common $sp^3$ diamond structure. The
properties of the model were determined using {\it ab initio}
total energy calculations. All significant experimental signatures
attributed to the platelets, namely, the lattice displacement
along the $[001]$ direction, the asymmetry between the $[110]$ and
the $[1\bar{1}0]$ directions, the infrared absorption peak
$B^\prime$ , and broad luminescence lines that indicate the
introduction of levels in the band gap, are naturally accounted
for in our model. The model is also very appealing from the point
of view of kinetics, since naturally occurring shearing processes
will lead to the formation of the metastable fault.

\end{abstract}


%\begin{multicols}{2}

\maketitle

Perhaps one of the oldest unanswered questions in diamond physics
concerns the nature and atomic structure of the extended defects
known as platelets. These $\{001\}$ planar defects were first
discovered more than sixty years ago \cite{Raman} by X-ray
diffraction experiments, which observed anomalous peaks
corresponding to $\{00h\}$ reflections. These were immediately
associated with lattice defects, since they are forbidden by
symmetry in a perfect diamond lattice. From then on, a plethora of
experimental data on this planar defect has been gathered, but a
complete understanding of its origin and microscopic structure is
still lacking. The current knowledge on this defect can be
summarized as follows: i) platelets have been detected only in
type Ia diamonds \cite{types}; ii) transmission electron
microscopy (TEM) experiments have shown that there is an asymmetry
between the $[110]$ and $[1\bar{1}0]$ directions \cite{Evans1};
iii) TEM experiments have also determined that platelets displace
the crystalline lattice by approximately $0.4 a_0$ \cite{Humble},
along the [001] direction, where $a_0$ is the lattice parameter of
diamond; iv) electron energy loss spectroscopy (EELS) experiments
indicate that the nitrogen content in platelets can vary from 6\%
to 61\% of a monolayer \cite{Bruley,Fallon,Kiflawi}; v) an
infrared absorption line around 1370 $\rm cm^{-1}$ is always
present in samples containing platelets \cite{Woods2}; vi)
platelets have been associated with broad luminescence bands, one
centered at 1.25 eV \cite{Desgreniers}, which reduces the
efficiency of optical windows made of natural diamond, and another
at 2.14 eV \cite{Collins}, and possibly with high energy
absorption and luminescence bands around 4.6 and 4.4 eV,
respectively \cite{Sobolev1}.

In the microscopic model first proposed for platelets in diamond,
Frank \cite{Frank} considered that the defect should be formed by
Si impurities replacing part of the carbon atoms in a $(001)$
plane. However, it was later shown that Si is an uncommon impurity
in diamond. Subsequently, it was demonstrated that diamonds type
Ia contain N impurities in the required concentration
\cite{Kaiser}. Following that, Lang \cite{Lang} proposed that the
defect would be constituted by a double layer of N atoms (one
layer of substitutional N atoms and one layer of interstitial N
atoms). However, as mentioned above, the N content in the
platelets can vary substantially \cite{Bruley,Fallon,Kiflawi},
ruling out N impurities as the main constituent of the defect.
Since foreign impurities could not account for the existence of
platelets, Evans \cite{Evans2} proposed that platelets could be
formed by interstitial carbon atoms. Based on Evans' idea, Humble
\cite{Humble} suggested that the double layer of nitrogen atoms in
Lang's model should be replaced by a double layer of carbon atoms,
thus establishing a so-called interstitial-aggregate model that
has since become the most accepted microscopic model for platelets
in diamond. More recently, Baker \cite{Baker} proposed another
model for platelets based on the aggregation of the $R1$ centers.

Very recently, several interstitial models, based on Humble's
proposal, have been extensively studied by Goss and co-workers
\cite{Goss}, using {\it ab initio} calculations. Their results
indicate that some of the experimental signatures of the platelets
can be explained by the interstitial model. However, the aggregate
of interstitials does not introduce states in the electronic band
gap, and hence it alone cannot account for the optical activity
experimentally observed. Goss \textit{et al.} propose that the
observed levels in the gap are due to vacancies and nitrogen
impurities segregated at the platelets. Therefore, one could
expect these properties to be somehow sample dependent. In fact,
the main difficulty with the interstitial aggregate model concerns
the energetics and the atomistic mechanism of its formation. The
formation energy of interstitial defects in the bulk is quite high
($\sim$ 12 eV). Goss {\it et al.} \cite{Goss} have proposed that
forming a Frenkel pair (a complex of an interstitial and a
vacancy) at the platelet, would result in a substantial reduction
in the interstitial formation energy.

Here we propose a new model for the microscopic structure of
platelets in diamond based on a entirely different mechanism. The
model does not require the creation of point defects, instead, the
defect is created by a shearing process of the crystal. It sprung
from the study of the $\gamma$-surface \cite{Vitek} or the
generalized stacking fault energy landscape for the $(001)$ slip
plane in diamond. In particular, we studied the behavior of the
$\gamma$-surface along the $[110]$ and $[1\bar{1}0]$ directions
using {\it ab initio} calculations. The model, aside from
accounting for all the known experimental signatures of platelets
in diamond, requires a much smaller activation energy for the
formation of the planar defect than the interstitial aggregate
model.

The results we present here were obtained through {\it ab initio}
calculations, within the framework of the density functional
theory (DFT) and the generalized gradient approximation (GGA),
using the SIESTA code \cite{siesta}. A double-zeta localized basis
set with polarization orbitals was used. The interaction between
the valence electrons and the ionic cores was modeled using {\it
ab initio} norm-conserving pseudopotentials. A 216-atom supercell
was used in most of the calculations, except for the vibrational
density of states calculations, where a 64-atom supercell was
employed. The $\Gamma$ point was used in the Brillouin zone
sampling for the 216-atom supercell, whereas for the calculations
using the 64-atom supercell 8 k-points were used. To check the
accuracy, we calculated the intrinsic stacking fault energy in the
$(111)$ slip plane, obtaining 0.279 $\rm J/m^2$, in excellent
agreement with the experimental value $0.285 \pm 0.040$ $\rm
J/m^2$ \cite{Suzuki}.

\begin{figure}[htbp]
\begin {center}
\includegraphics[angle = 270,width = 8.0 cm]{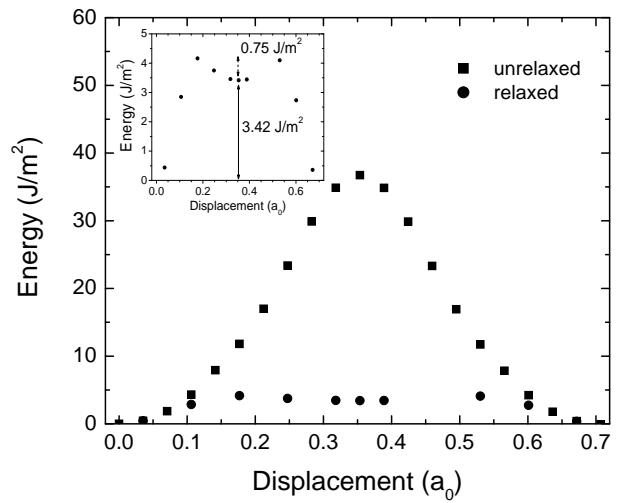}
\caption{Profile of the $\gamma$-surface of the $(011)$ plane
along the $[1\bar{1}0]$. The inset shows the details of the energy
minimum for the displacement $(a_0/4)\langle1\bar{1}0\rangle$
after atomic relaxation.} \label{fig1}
\end{center}
\end{figure}

The $\gamma$-surface for a given crystalline plane is obtained by
cutting the infinite crystal in half along the plane and shearing
the upper part with respect to the lower part by a displacement
(slip) vector, which belongs to the plane. Our results for the cut
of the $\gamma$-surface along the $[1\bar{1}0]$ direction are
shown in Fig. 1. One of the curves displays the unrelaxed
$\gamma$-surface, which exhibits a maximum corresponding to a
displacement of $(a_0/4)[1\bar{1}0]$. For this displacement, the
atoms in both sides of the cut approach each other very closely
and are aligned along the $[001]$ direction. Fig. 1 shows a quite
large release of energy that results from allowing the atoms to
relax along the $[001]$ direction only. In fact, we found that the
slip at $(a_0/4)[1\bar{1}0]$ corresponds to a local minimum of the
$\gamma$-surface along this direction, in contrast with the case
of Si and Ge, where this configuration corresponds to the
so-called unstable stacking fault. On the other hand, shearing
along the $[110]$ direction does not bring the system to any
intermediate stable configuration, and our calculations confirm
that the $\gamma$-surface does not exhibit any local minimum, a
behavior in this case similar to what is found in silicon and
germanium \cite{Juan}. Since at first we allowed relaxation of the
$(a_0/4)[1\bar{1}0]$ configurations along the $[001]$ direction
only, to verify that this is a local minimum of the
$\gamma$-surface we allowed all the atoms to move freely. We found
that, indeed, the atomic structure is stable. That such
configuration is stable in diamond and not in silicon and
germanium, stems from the stability of the $sp^2$ bonding in
diamond.

\begin{figure}[htbp]
\begin {center}
\includegraphics[width = 7.4 cm]{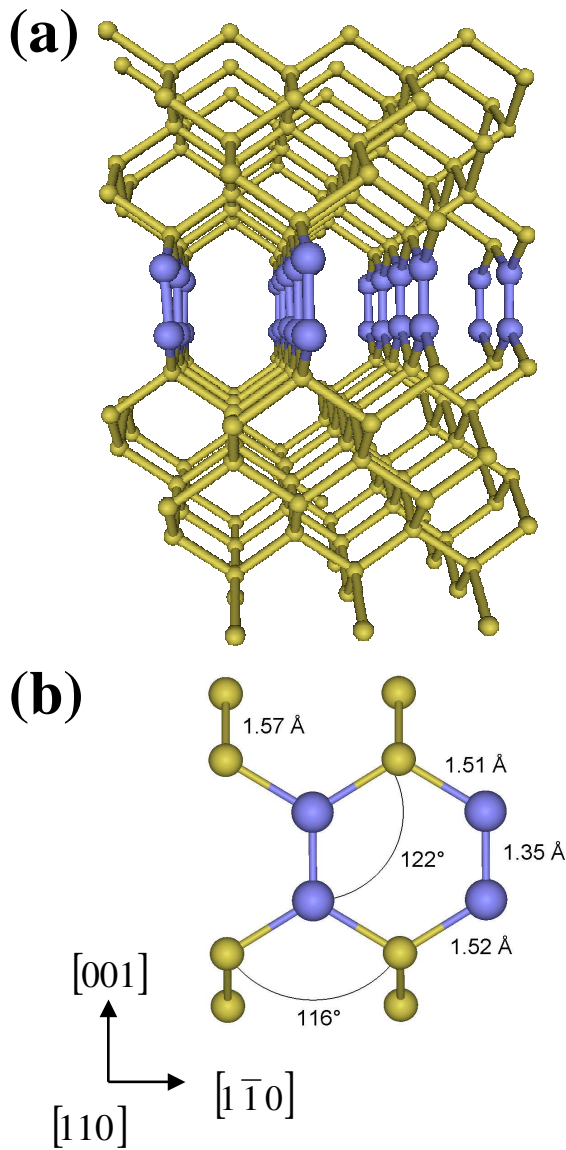}
\caption{(a) Atomic structure of the planar defect. The blue
colored atoms are three-fold coordinated carbon atoms and the gold
colored atoms are four-fold coordinated carbon atoms, (b) Detail
of the core of the planar defect indicating bond lengths and
angles.} \label{fig2}
\end{center}
\end{figure}

Fig. 2a depicts the relaxed atomic structure corresponding to the
metastable stacking fault at $(a_0/4)[1\bar{1}0]$. The blue
colored carbon atoms are three-fold coordinated, whereas the gold
colored ones are the usual four-fold coordinated carbon atoms in
diamond. The core of the defect is formed by a double layer of
three-fold coordinated carbon atoms. In Fig.~2(b), we display the
structure of one unit of the core of the defect (with a few
neighboring atoms), when viewed along the $[110]$ direction. This
unit is formed by the repetition of a hexagonal ring consisting of
four three-fold coordinated atoms and two four-fold coordinated
atoms. The length of the bonds between three-fold coordinated
atoms is 1.35 $\textrm{\AA}$, which is even shorter than the
carbon bonds in graphite (1.42 $\textrm{\AA}$), while the bonds
between three-fold coordinated and four-fold coordinated atoms are
about 1.52 $\textrm{\AA}$ long, very close to the value of 1.57
$\textrm{\AA}$ we obtained for the bond length in bulk diamond.
Note that, in Fig.~2(b), the length of the bonds between the
four-fold coordinated atoms just outside the defect core is
already the bond length in bulk diamond. Therefore, this structure
can be seen as two semi-infinite pieces of diamond "glued"
together by chemical bonds that are 1.35 $\textrm{\AA}$ long,
which is equal to $0.38 a_0$, basically the lattice displacement
observed in platelets. Moreover, it is also easy to see that the
atomic structure of the core along the $[1\bar{1}0]$ direction is
different from that along $[110]$ direction, reflecting the
asymmetry of the $\gamma$-surface between the two directions. This
extended defect has been discussed before \cite{Gruen}, however,
within a completely different context, related to grain boundaries
in artificially grown diamonds.

\begin{figure}[htbp]
\begin {center}
\includegraphics[angle = 270,width = 8.0 cm]{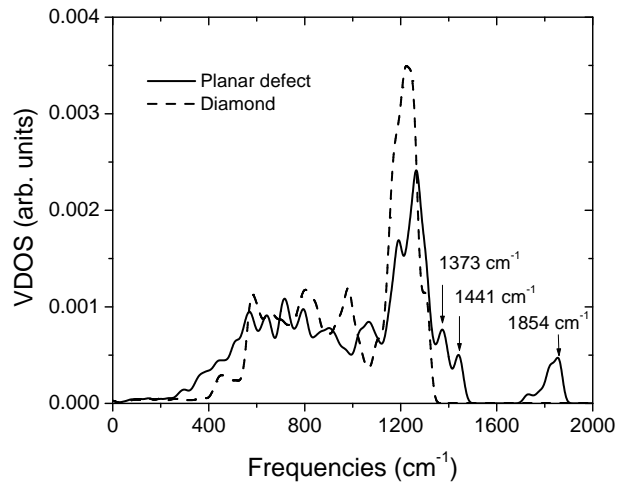}
\caption{Vibrational density of states in bulk diamond and in
diamond containing a planar defect} \label{fig3}
\end{center}
\end{figure}

In Fig. 3, we show our \textit{ab initio} results for the
vibrational density of states (VDOS) for the planar defect and for
bulk diamond. First, we remark that our calculations can reproduce
quite well the bulk VDOS.  Moreover, our results show clearly the
appearance of three peaks above the highest allowed frequency in
the bulk. Two of these peaks lie just above the bulk band edge, at
1373~$\rm cm^{-1}$ and 1441~$\rm cm^{-1}$, values which are in
very good correspondence with the frequencies (1372~$\rm cm^{-1}$
and 1426~$\rm cm^{-1}$) of the experimental bands usually
associated with platelets\cite{Woods2}. The first peak at
1373~$\rm cm^{-1}$ is more intense and can be associated with the
$B^\prime$ band. By computing the vibration frequency of the
stretch mode of the compressed bonds in the core of the defect, we
determined that the third peak in Fig.~3, at $1854~\rm cm^{-1}$,
is a localized mode associated with these compressed $sp^2$ bonds
in the core~\cite{Gironcoli}. As far as we know this frequency is
out of the range usually investigated in infrared absorption
studies of platelets, 900~$\rm cm^{-1}$ to 1650~$\rm cm^{-1}$
\cite{Woods3}. This is quite interesting, because this third peak
opens the possibility that the validity of our model could be
experimentally investigated.

\begin{figure}[htbp]
\begin {center}
\includegraphics[angle = 270,width = 8.0 cm]{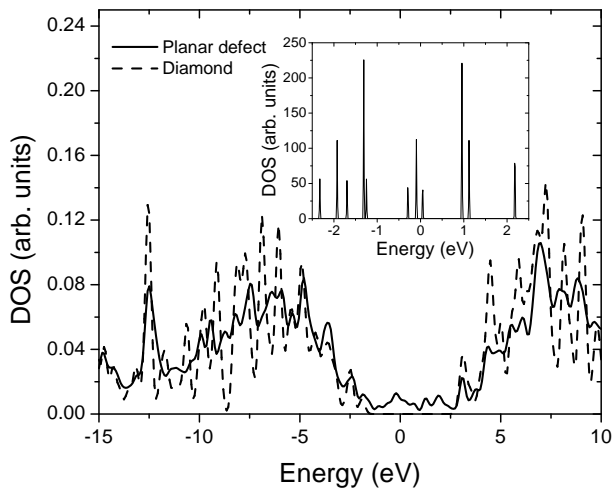}
\caption{Electronic density of states of bulk diamond and of
diamond containing a planar defect. The inset shows the levels
without broadening. } \label{fig4}
\end{center}
\end{figure}

We now discuss the electronic structure of the metastable fault.
Fig. 4 depicts both the electronic density of states of a bulk
diamond cell and of a supercell containing the planar defect. In
the figure, the zero in the energy scale corresponds to the
highest occupied level. The first important point to note is that
the planar defect gives rise to deep levels in the band gap. The
inset in Fig. 4 shows the levels in the diamond band gap without
broadening. The lowest empty levels lie about 1.0~$\rm eV$. Also,
in the inset, one can see empty levels about 2.1~eV. Above that,
there are empty levels that coincide with the bottom of the
conduction band of the bulk at about 3.1~eV (not shown in the
inset). Our results agree with the luminescence bands at 1.25~$\rm
eV$ and 2.14~eV, usually associated with the platelets. It is
important to emphasize that these levels appear in the gap as
consequence of the planar defect, without the intervention of
either native defects or impurities.

We have also investigated the possibility of the occurrence of a
similar intermediate metastable stacking fault in the $(111)$ slip
plane. For this plane, a very similar situation, with the atoms
adjacent to the cut getting near each other, also happens when the
system is sheared along the $[1\bar{2}1]$ direction by a
displacement of $(\sqrt{6}/12) a_{0}$. In the case of Si and Ge,
previous calculations \cite{Juan} indicate that the atomic
configuration obtained after relaxation is still a distinct local
maximum for the $\gamma$-surface of the glide set. In the case of
diamond, the $\gamma$-surface resembles a plateau, which turned
out to be unstable, with the atomic planes adjacent to the cut
always slipping to either the perfect crystal or the intrinsic
stacking fault configurations. Since the atomic density in the
$(111)$ plane is larger than in the $(001)$ plane, the $sp^2$
hybridization of the atoms adjacent to the cut should be less
favorable than the $sp^3$, which would explain why platelets are
not observed in the $(111)$ plane.

We now compare our model with the self-interstitial model. First,
according to the results in Fig.~1, the activation energy per unit
of area required to form the metastable stacking fault in our
model is (3.42+0.75)~$\rm J/m^2$ = 0.26~$\rm eV/\AA^{2}$. This
activation energy per unit of area should be compared with the
activation energy per unit of area required in the interstitial
model. According to Goss {\it et al.}, the lowest energy to create
a Frenkel pair at the platelet, for climb of the jog, and to
diffuse the vacancy away from the platelet is about 8.8~$\rm eV$,
which would imply in an energy per unit of area of 1.39~$\rm
eV/\AA^{2}$ \cite{Area}. Therefore, the activation energy for the
formation of the stacking fault in our model is about 5 times
smaller than the corresponding activation energy in the model
based on interstitials. Furthermore, we have performed preliminary
calculations which indicate that the presence of nitrogen
aggregates can further reduce the activation energy for the
formation of the metastable stacking fault. The physical reason
for this is the fact that since N is a trivalent impurity in
diamond, less bonds have to be broken by the shearing process that
creates the defect. Aside from a more favorable energetics of
formation, we believe that our model accounts for the optical
properties of platelets in a more complete and straightforward
way. Also, regarding symmetry properties, the study by Goss {\it
et al.} indicates that the interstitials can be arranged in a
number of different structures, which have similar formation
energy, not all of them exhibiting the asymmetry between the
$[110]$ and the $[1\bar{1}0]$ directions. In contrast, this
asymmetry is essentially built-in in our model. In a nutshell, we
propose a new model for platelets in diamond based on the shearing
process of $\{001\}$ planes that explains all experimental data
available.

\begin{acknowledgments}

We acknowledge the financial support by the Brazilian funding
agencies FAPESP, CNPq, and FAEP-UNICAMP; and the computer
resources granted by CENAPAD-SP.

\end{acknowledgments}

%\pagebreak

%\pagebreak

%\end{references}
%\end{multicols}
%\pagebreak


\begin{thebibliography}{99}
\bibitem{Area} In the interstitial model, the interstitials
are arranged in a square lattice whose lattice constant is
$(\sqrt{2}/2)a_{0}$.
\end{thebibliography}
 \end{document}